\documentclass[journal]{IEEEtran}
\IEEEoverridecommandlockouts
\usepackage{cite}
\usepackage{amsmath,amssymb,amsfonts}
\usepackage{algorithm}
\usepackage{algpseudocode}
\usepackage{graphicx}
\usepackage{textcomp}
\usepackage{xcolor}
\usepackage{url}
\usepackage{tikz}
\usepackage{hyperref} 
\usepackage{pgfplots}
\pgfplotsset{compat=1.18}
\def\BibTeX{{\rm B\kern-.05em{\sc i\kern-.025em b}\kern-.08em
    T\kern-.1667em\lower.7ex\hbox{E}\kern-.125emX}}
\begin{document}

\title{A Novel Short-Term Anomaly Prediction for IIoT with Software Defined Twin Network\\
}

\author{
\IEEEauthorblockN{Bilal Dalgıç, Betül Şen and Müge Erel-Özçevik}
\IEEEauthorblockA{\textit{Department of Software Engineering} \\
\textit{Manisa Celal Bayar University}, Manisa, Turkey \\
Emails: 222803043@ogr.cbu.edu.tr, 222803014@ogr.cbu.edu.tr, muge.ozcevik@cbu.edu.tr}
}
\markboth{Accepted by 2025 IEEE Globecom Workshops-TwinNetApp, ©2025 IEEE}{}
\maketitle

\begin{abstract}

Secure monitoring and dynamic control in an IIoT environment are major requirements for current development goals. We believe that dynamic, secure monitoring of the IIoT environment can be achieved through integration with the Software-Defined Network (SDN) and Digital Twin (DT) paradigms. The current literature lacks implementation details for SDN-based DT and time-aware intelligent model training for short-term anomaly detection against IIoT threats. Therefore, we have proposed a novel framework for short-term anomaly detection that uses an SDN-based DT. Using a comprehensive dataset, time-aware labeling of features, and a comprehensive evaluation of various machine learning models, we propose a novel SD-TWIN-based anomaly detection algorithm. According to the performance of a new real-time SD-TWIN deployment, the GPU-
accelerated LightGBM model is particularly effective, achieving a balance of high recall and strong classification performance.
\end{abstract}

\begin{IEEEkeywords}
Software Defined Networks, Digital Twin, Anomaly Prediction, Industrial IoT, Machine Learning
\end{IEEEkeywords}
\vspace{-1em}
\section{Introduction}
The rise of Industry 5.0 has spurred the integration of Cyber-Physical Systems (CPS) and the Industrial Internet of Things (IIoT), creating highly interconnected and intelligent environments \cite{jagatheesaperumal2023building}. According to Ericsson's report, secure monitoring and dynamic control of the network is one of the major point of such network \cite{Ericsson}. To handle this requirement in low OPEX/CAPEX, the current tenants deploy Software-Defined Network (SDN) based topology control. It decouples data and control planes and enables to easy implementation of new network services via northbound interface of controller \cite{Muge}. Early SDN-based orchestration for multi-RAT offloading showed that a centralized controller with a global network view can improve end-to-end QoS and overall user satisfaction \cite{arslan2014sdoff}. 

On the other hand; at the heart of the secure transformation of IIoT lies the Digital Twin (DT) paradigm: a virtual, high-fidelity replica of a physical asset or system that is continuously updated with real-world data. This dynamic mirroring enables advanced simulation, monitoring, and optimization, promising unprecedented efficiency and resilience. However, the performance of a DT is fundamentally constrained by its ability to maintain a timely and accurate state synchronization with its physical counterpart. As analyzed by Li et al. \cite{li2023performance}, this synchronization involves a delicate trade-off between timeliness, data distortion, and energy sustainability, where significant delays can render the twin ineffective for real-time applications.

This tight coupling between the physical and digital worlds introduces significant security vulnerabilities in IIoT environment. Traditional security mechanisms often fall short, as they are typically reactive and struggle to adapt to the evolving threat landscape. While recent advancements push the boundaries of real-time detection \cite{alhawawreh2024digital, yigit2025digital} and threat analysis \cite{yigit2023twinpot}, a truly proactive security posture requires the ability to predict threats before they manifest. %This paper bridges this critical gap by introducing a framework for short-term anomaly prediction at \textit{T\,+15 s}, which forecasts whether an incident will occur within the next 15 seconds and thereby shifts the paradigm from reaction to pre-emption.
\vspace{-0.5em}
\subsection{Related Work and Contributions}
The application of DTs in cybersecurity has evolved from passive monitoring to active defense. Yigit \textit{et al.}~\cite{yigit2025digital} demonstrated the viability of lightweight, DT-enabled intrusion detection in resource-constrained edge networks, achieving $>99\%$ detection accuracy with a latency of $\approx\!0.22\,\text{ms}$, while their “TwinPot” honeypot~\cite{yigit2023twinpot} showed how high-fidelity DTs can safely analyse attacker behaviour. Addressing the challenge of evolving threats, Al-Hawawreh and Hossain \cite{alhawawreh2024digital} proposed a DT framework with online ensemble learning to continuously adapt to evolving attack patterns. These systems, however, remain focused on detecting ongoing attacks.

The effectiveness of any real-time DT application hinges on robust "virtual-real synchronization." Foundational work by Li et al. \cite{li2023performance} provides a formal analysis of the trade-offs between timeliness, distortion, and sustainability in DT networks, highlighting the inherent performance limits. On a broader scale, the vision for DTs in Industry 5.0, as outlined by Jagatheesaperumal and Rahouti \cite{jagatheesaperumal2023building}, positions them as central enablers for resilient, human-centric industrial operations. 

Therefore, we've belived that these challenges can be handle by SDN based DT with an intelligent anomaly detection framework for IIoT network. Then, we've proposed SD-TWIN based topology control and monitoring. Our work builds on the aforementioned DT concepts, applying them to a different problem. SDN makes this possible thanks to its easy implementation and configuration environment for new network services. The core of our proposed framework is built on principles of time-series forecasting, drawing from a rich body of literature surveyed by Benidis et al. \cite{benidis2022deep}. Unlike unsupervised, reactive methods such as those proposed by Singh and Srivastav \cite{singh2021novel}, which identify existing anomalies, our approach uses supervised learning on time-aware labels to predict future incidents. In light of these challenges and opportunities, this paper makes the following key contributions:
\begin{itemize}
\item A novel topology control via SDN based DT,
    \item A novel framework for short-term anomaly prediction, enabling a DT to anticipate threats,
    \item A novel dataset construction and a time-aware labeling strategy that transforms standard network security datasets for predictive learning tasks,
    \item A comprehensive evaluation of various machine learning models, analyzing the trade-offs between predictive accuracy, computational cost, and suitability for real-time DT deployment.
\end{itemize}
%The remainder of this paper details our system architecture, dataset preprocessing, model design, and experimental results, concluding with a discussion on future work toward fully autonomous, predictive cyber-defense systems.
%\vspace{-0.5em}
\section{Proposed System Architecture}
\vspace{-1.1em}
\begin{figure}[htbp]
  \centering
  \includegraphics[width=\linewidth]{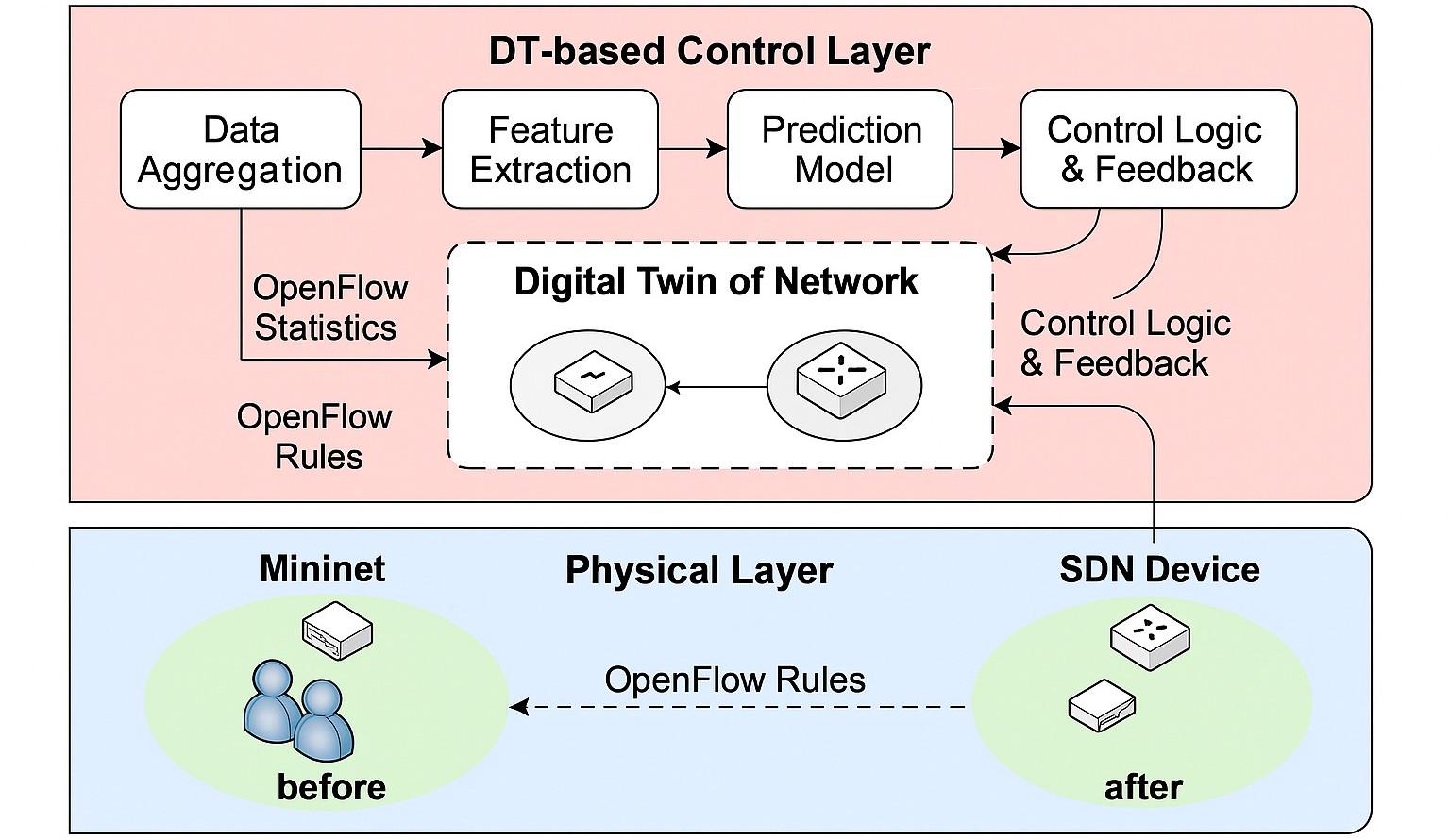}
  \vspace{-2em}
  \caption{Four-layer SDN based digital-twin pipeline.}
  \label{fig:overall-arch}
  \vspace{-0.7em}
\end{figure}

The proposed system architecture of SD-TWIN is given in Figure~\ref{fig:overall-arch}. It depicts the end-to-end digital-twin pipeline that collects flow statistics from a Mininet-emulated IIoT/enterprise topology via the SDN controller's interface, i.e. \textit{ONOS REST API}. This architectural pipeline, where a centralized entity gathers distributed network telemetry for predictive analysis and proactive control, is a powerful paradigm. It has been successfully applied in different layers of the network stack, such as for managing physical-layer interference by collecting signal data to preemptively optimize resources \cite{GutierrezEstevez2012SpatioTemporal}. It trains the data aggeragated at DT period of T, and it results T+15 anomaly-prediction model offline. It delivers real-time predictions to administrators, and the SD-TWIN model consists of four layers:

\begin{itemize}
\item \emph{Physical Layer:}

We deploy representative IIoT sensors, PLCs and enterprise hosts inside
a Mininet virtual testbed.  Mininet provides fine-grained control over
link delays and bandwidth, enabling realistic traffic generation
scenarios for both \textit{benign} and \textit{attack} flows.

\item \emph{SDN \& Telemetry Layer (ONOS API):}

The emulated switches run the OpenFlow protocol and register with the
ONOS controller.  Every 5s, ONOS exposes per-switch and per-flow
counters—\texttt{flow\_count}, \texttt{total\_packets},
\texttt{total\_bytes}, \texttt{avg\_packet\_size}, and
\texttt{link\_count}—through its REST API; these JSON payloads are
streamed to a Kafka topic for downstream analytics.  The controller can
also \emph{enforce} mitigation rules (e.g.\ drop, rate-limit, reroute)
whenever the twin flags high-risk flows.

\item \emph{Digital-Twin \& Analytics Layer:}

A LightGBM-GPU model is trained offline on historical Mininet + ONOS
flows that have been labelled with the temporal
\texttt{label\_t+15}.  After training, the model is serialised and loaded into a
FastAPI micro-service that performs real-time inference.  Incoming
feature vectors are normalised, passed to the model, and the predicted
risk score~$p\!\in\![0,1]$ is published to Redis Streams.

\paragraph*{Self-Updating Twins}  
The micro-service continuously appends new flow records and their
ground-truth labels (derived from attack logs) to a training buffer.
Once the buffer reaches a threshold (e.g.\ 100 k records), an automatic
re-training job is triggered, and the updated model is hot-swapped with
zero downtime.

\paragraph*{Hierarchical Twin Federation}  
Edge-level surrogate models—pruned versions of the main LightGBM—run on
resource-constrained Raspberry Pi gateways inside the Mininet network.
These edge twins handle latency-sensitive decisions locally, while
periodically synchronising parameters with the central twin through
gRPC.

\paragraph*{Explainability and Trustworthiness}  
Each prediction is accompanied by SHAP feature attributions that are
pushed to the dashboard and stored for forensic auditing.

\item \emph{Visualisation \& Control Layer:}
\vspace{-1em}

\begin{figure}[htbp]
  \centering
  \includegraphics[width=\columnwidth]{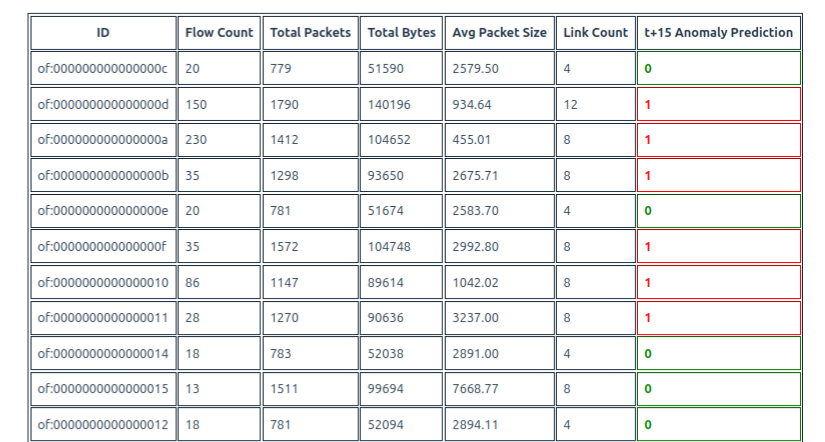}
  \vspace{-1.5em}
  \caption{Digital-twin control panel with T+15 anomaly predictions.}
  \label{fig:twin-ui}
  \vspace{-0.1em}
\end{figure}

A React + Cytoscape.js front-end renders the live network topology, while a real-time \textbf{device table} at the right side of the screen
(Figure ~\ref{fig:twin-ui}) displays flow statistics and the binary T\,+15 prediction for every switch or host.  
The table columns are:

\begin{itemize}
  \item \textbf{ID} – OpenFlow switch/host identifier  
        (e.g.\ \texttt{of:000000000000000c});
  \item \textbf{Flow Count} – number of active flows currently handled by the device;
  \item \textbf{Total Packets} – cumulative packet count for those flows;
  \item \textbf{Total Bytes} – cumulative byte count;
  \item \textbf{Avg Packet Size} – mean bytes per packet;
  \item \textbf{Link Count} – degree of the device in the topology graph;
  \item \textbf{t\,+15 Anomaly Prediction} – binary output 0 (benign) or 1 (anomaly expected within 15 s).
\end{itemize}
\end{itemize}
In Figure  \ref{fig:twin-ui}, Clicking a row that shows an anomaly (\textbf{1}) opens a modal with SHAP feature attributions; a Mitigate button lets the operator push an immediate flow-rule patch (drop, rate-limit, reroute) to ONOS. As a softer first-line response before hard isolation, the controller can perform two-scale AI-assisted CST retuning—local at STAs and global at APs—to balance throughput, collisions, hidden/exposed terminals, and fairness under forecasted contention \cite{ak2020fsc}.

\begin{algorithm}[htp]
\caption{Operational flow of SD-TWIN}\label{alg:sd_twin}
\begin{algorithmic}[1]
\Require Flow-level traffic records, labeled datasets
\Ensure Real-time $t{+}15$ anomaly forecasts with SDN controller actions
\State Collect flow statistics from ONOS
\State Preprocess features and assign temporal labels (\texttt{label\_t+15})
\State Train candidate models (RF, RF+GS, MLP, LGBM, LGBM-GPU, DL) and select best-performing
\State Initialize digital twin with selected model
\While{system operational}
  \State Capture current flow metrics and predict anomaly at $t{+}15$
  \If{anomaly predicted}
    \State Send control command to ONOS for mitigation
  \EndIf
\EndWhile
\State Update twin state and periodically retrain with new data
\end{algorithmic}
\end{algorithm}

The operational flow of the SD-TWIN framework is summarized in Algorithm ~\ref{alg:sd_twin}, encompassing data preprocessing, anomaly labeling, model development, and controller-level threat mitigation. The details of data aggreagtion and model design are given in the following section.
%\vspace{-1em}

\section{Model Design and Training}
\subsection{Dataset Description}
In this study, we utilize two publicly available, well-documented benchmark datasets—\textbf{CICAPT-IIoT2024} and \textbf{CIC-IDS2017}—to evaluate the performance and generalizability of our proposed anomaly prediction models across different network environments. While CICAPT-IIoT2024 focuses on Industrial IoT (IIoT) infrastructure and Advanced Persistent Threats (APT), CIC-IDS2017 represents a more traditional enterprise network with a broader spectrum of contemporary cyberattacks.

\subsubsection{CICAPT-IIoT2024 Dataset}
The CICAPT-IIoT2024 dataset \cite{ghiasvand2024cicapt, ghiasvand2024resilience}, developed by the Canadian Institute for Cybersecurity, provides temporally structured APT-related data captured in a realistic IIoT testbed. This testbed simulates industrial control systems using two Raspberry Pi devices, Ubuntu and Kali Linux virtual machines, and an NS-3 network simulator. The dataset includes provenance and network traffic logs, along with attack metadata. It covers over 20 APT techniques across 8 major MITRE ATT\&CK tactics, inspired by known adversarial groups. Our experiments focus on network traffic, extracting features like \texttt{flow\_count}, \texttt{total\_packets}, and \texttt{link\_count}. A binary temporal label \texttt{label\_t+15} is generated to indicate anomalies within 15 seconds after the current record, facilitating short-term predictive anomaly detection.

\subsubsection{CIC-IDS2017 Dataset}
The CIC-IDS2017 dataset \cite{sharafaldin2018toward}, also from the Canadian Institute for Cybersecurity, captures five days of realistic enterprise network activity, including both benign traffic and various cyberattacks. The test environment features a comprehensive enterprise network topology with switches, routers, firewalls, and systems running Windows, Linux, and macOS, supporting 25 virtual users and real-world attack nodes. It includes various attack types such as Brute Force, DoS, Botnet, Infiltration, Heartbleed, Web attacks, and Port scanning. Traffic data is converted to labeled flow-based CSV files using CICFlowMeter, providing over 80 features. Similar to CICAPT-IIoT2024, a temporal binary label \texttt{label\_t+15} is created for each record, enabling the model to anticipate imminent threats. This dataset is crucial for evaluating model robustness across heterogeneous IT environments.

\subsection{Data Preprocessing and Labeling}

To enable proactive anomaly prediction aligned with a digital twin paradigm, both datasets underwent tailored preprocessing pipelines and temporal labeling strategies. Each record was transformed to support T+15 anomaly forecasting, i.e., predicting whether an anomaly would occur within 15 seconds following a given flow entry. Below we detail the dataset-specific preprocessing steps.

\subsubsection{CICAPT-IIoT2024 Preprocessing}

The CICAPT-IIoT2024 dataset was first sorted chronologically using the \texttt{ts} (timestamp) field to ensure temporal coherence. A binary label \texttt{label\_t+15} was generated for each row to indicate whether any attack occurs within the subsequent 15-second window. This was achieved by performing a forward-index search using a binary search mechanism (\texttt{np.searchsorted}), and checking for the presence of any anomalies between the current timestamp and the future index. 

After label generation, features irrelevant to short-term anomaly dynamics—such as source and destination IP addresses, low-variance or redundant attributes, and original attack labels—were removed to reduce dimensionality and prevent data leakage. Remaining categorical attributes were label-encoded, and the dataset was split into training and testing subsets using stratified sampling to preserve class distributions.

\subsubsection{CIC-IDS2017 Preprocessing}

The CIC-IDS2017 dataset consists of flow-based features derived from labeled PCAPs using CICFlowMeter. In this study, 18 traffic and timing-based attributes such as \texttt{Flow Bytes/s}, \texttt{Flow IAT Mean}, and \texttt{Packet Length Std} were selected as core features. To preserve contextual integrity, IP addresses and port numbers were retained, while highly redundant or purely categorical features were excluded.

As in CICAPT-IIoT2024, a temporal binary label \texttt{label\_t+15} was created for each record, indicating whether an attack occurs within 15 seconds after the current flow. This allowed the model not only to detect present threats but also to anticipate imminent ones. The final dataset structure thus supports short-term predictive learning and aligns with the real-time decision-making requirements of digital twin architectures.

\subsection{SD-TWIN Models}
A diverse set of machine–learning and deep-learning models was trained on the two benchmark datasets in order to study how architectural choices and imbalance-handling techniques affect short-term (T\,+15) anomaly forecasting.

\subsubsection{CICIDS2017 Models}

For the enterprise-network scenario represented by CICIDS2017, three models were considered:

\begin{itemize}
    \item \textbf{RF}\,: A baseline Random Forest trained on Standard\-Scaler–normalised features with fixed 
        hyper-parameters (\texttt{n\_estimators}=50, \texttt{max\_depth}=10).  
        No hyper-parameter optimisation was performed, which kept the training time short and provided a
        reference point for evaluating the benefits of more advanced tuning strategies.

    \item \textbf{RF\,+\,GS}\,: A Random Forest whose hyper-parameters were tuned with \texttt{GridSearchCV}. 
        Multiple combinations of \texttt{n\_estimators}, \texttt{max\_depth}, and \texttt{min\_samples\_split} 
        were evaluated under a 3-fold stratified cross-validation scheme.  
        Although the exhaustive search noticeably prolonged training, it produced a configuration that 
        delivered the highest validation accuracy and F1/F2 scores among the CICIDS2017 candidates, 
        confirming improved generalisation over the default RF baseline.

  \item \textbf{MLP}\,: A fully connected Multi-Layer Perceptron. Input features were normalised with \texttt{MinMaxScaler}; the network was trained with the Adam optimiser, early stopping, and a patience of ten epochs to curb over-fitting.
\end{itemize}

\subsubsection{CICAPT-IIoT2024 Models}

The IIoT-oriented CICAPT-IIoT2024 dataset exhibited a pronounced class imbalance, for which Random UnderSampling (RUS) was applied before model fitting. Three variants were evaluated:
\begin{itemize} 
  \item \textbf{LGBM}\,: A gradient-boosted LightGBM classifier trained on the RUS-balanced set, with categorical features label-encoded, 300 boosting rounds, and class-balanced loss weights. The decision threshold was chosen from the precision–recall curve as the earliest point meeting \emph{Precision}~$>0.20$ and \emph{Recall}~$>0.50$, thereby aligning the model’s operating point with the recall-centric objectives of T\,+15 anomaly prediction.

  \item \textbf{LGBM-GPU}\,: A GPU-accelerated LightGBM that additionally incorporated feature-selection heuristics and decision-threshold optimisation. After training, the precision–recall curve was scanned to identify the threshold that maximised the F\textsubscript{2}-score, thereby favouring recall—a critical property for early-warning systems.
   \item \textbf{DL}\,: A fully connected neural network trained on the RUS-balanced, standard-scaled feature set.  
        The architecture comprises two hidden layers of 128 and 64 ReLU units, each followed by 0.30 dropout for
        regularisation.  Training used the Adam optimiser, a batch size of 512, and early stopping (patience = 5)
        over a maximum of 30 epochs.  After inference, 100 candidate thresholds (0.05–0.95) on the
        precision–recall curve were scanned, and the value maximising the F\textsubscript{2}-score was selected for
        final classification.
\end{itemize}
%\vspace{-3em}
\section{Performance Evaluation and Results}
%\vspace{-3em}
To provide a more informative evaluation than raw accuracy on highly imbalanced data, we report \textit{precision}, \textit{recall}, the recall-weighted $F_{2}$‐score, and the area under the ROC curve (AUC).  
Model predictions are first decomposed into four fundamental outcomes:
\begin{itemize}
  \item \textbf{True Positive (TP)} – correctly classifying a real attack instance as \emph{Attack};
   \vspace{0.2em}
  \item \textbf{False Positive (FP)} – mistakenly classifying benign traffic as \emph{Attack};
  \vspace{0.2em}
  \item \textbf{False Negative (FN)} – missing a real attack instance (\emph{Attack} predicted as \emph{Benign});
  \vspace{0.2em}
  \item \textbf{True Negative (TN)} – correctly recognising benign traffic as \emph{Benign}.
\end{itemize}
By using these definitions, the following performance evaluation metrics are calcualted as follows:
\begin{align}
\text{Precision} &= \frac{TP}{TP + FP}\\[4pt]
\text{Recall}    &= \frac{TP}{TP + FN}\\[4pt]
F_{2}            &= \frac{5\,\text{Precision}\,\text{Recall}}
                        {4\,\text{Precision} + \text{Recall}}\\[4pt]
\text{AUC}       &= \int_{0}^{1} \text{TPR}(\text{FPR})\, d(\text{FPR})
\end{align}
where $F_{2}$ (with $\beta=2$) weights recall four times more heavily than precision — a desirable choice in proactive anomaly detection, where overlooking an attack is costlier than issuing a false alarm.

\begin{table}[htp]
\caption{Key Evaluation Metrics for CIC-IDS2017 and CICAPT-IIoT}
\centering
\scriptsize
\begin{tabular}{|l|l|c|c|c|c|r|}
\hline
\textbf{Dataset} & \textbf{Model} & \textbf{Prec.} & \textbf{Rec.} & \textbf{F2} & \textbf{AUC} & \textbf{Time (s)} \\
\hline
CIC-IDS & RF & 0.868 & 0.918 & 0.904 & 0.960 & 90.0 \\
CIC-IDS & RF+GS & 0.890 & 0.920 & 0.912 & 0.960 & 2906 \\
CIC-IDS & MLP & 0.863 & 0.866 & 0.913 & 0.961 & 709 \\ 
CICAPT & LGBM & 0.200 & 0.996 & 0.560 & 0.997 & 24 \\
CICAPT & LGBM-GPU & 0.615 & 0.952 & 0.822 & \textbf{0.998} & 5631 \\
CICAPT & DL & 0.025 & 0.193 & 0.083 & 0.734 & 95 \\
\hline
\end{tabular}
\label{tab:key_metrics}
%\vspace{-1em}
\end{table}

Table~\ref{tab:key_metrics} summarizes the performance of selected models across CIC-IDS2017 and CICAPT-IIoT datasets. Among all models, the GPU-accelerated LightGBM model (LGBM-GPU) achieved the best results on the CICAPT-IIoT dataset, with the highest F2-score (\textbf{0.822}) and ROC AUC (\textbf{0.9982}). These values indicate both high anomaly recall and excellent classification ability, which are critical for proactive T+15 anomaly prediction. In contrast, deep learning models such as DL + RUS yielded significantly lower precision and recall, despite faster training times. On the CIC-IDS2017 dataset, Random Forest models achieved robust and balanced performance, with the grid search variant reaching a recall of 0.92 and F2-score of 0.9116. These findings highlight the importance of combining recall-oriented scoring (F2) with ROC-based discrimination when evaluating models for time-aware cybersecurity prediction.
\vspace{1 em}

%\vspace{-0.8em}
\begin{figure}[h]
\centering
\begin{tikzpicture}
\begin{axis}[
    width=0.9\linewidth,
    height=4cm,
    xlabel={Model},
    ylabel={Time (s)},
    symbolic x coords={RF, RF+GS, MLP, LGBM, LGBM-GPU, DL},
    xtick=data,
    xticklabel style={rotate=30, anchor=east, font=\scriptsize},
    ymajorgrids,
    grid style=dashed,
    enlargelimits=0.1,
    bar width=8pt,
    ymin=0,
    ymax=6200,   % <-- burası eklendi
    yticklabel style={
        /pgf/number format/.cd,
        1000 sep={},
        precision=0
    },
    nodes near coords,
    every node near coord/.append style={font=\tiny},
    nodes near coords align={vertical},
]
\addplot[ybar,fill=gray!50] coordinates {
  (RF,90) (RF+GS,2906) (MLP,709) (LGBM,24) (LGBM-GPU,5631) (DL,95)
};
\end{axis}
\end{tikzpicture}
\vspace{-0.8em}
\caption{Training time of models on CIC-IDS2017 and CICAPT-IIoT.}
\label{fig:training-time}
\vspace{-1em}
\end{figure}
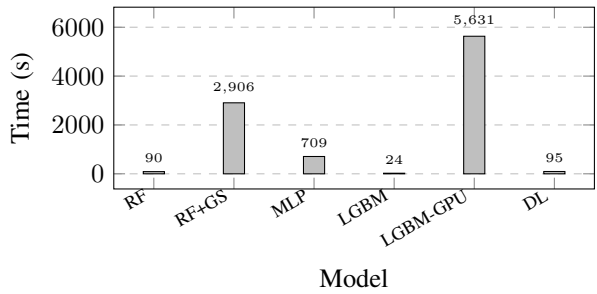

Figure~\ref{fig:training-time} complements these accuracy-centric results by illustrating the
computational cost of each candidate: while the GPU-accelerated LGBM reaches the highest
F\textsubscript{2} and AUC, its training time (5\,631 s) is **two orders of magnitude** longer than
the vanilla LGBM, which converges in just 24 s. Such a gap is critical for digital-twin
deployments that require frequent retraining or on-device updates.

\vspace{-1em}
\begin{figure}[h]
\centering
\includegraphics[width=\columnwidth]{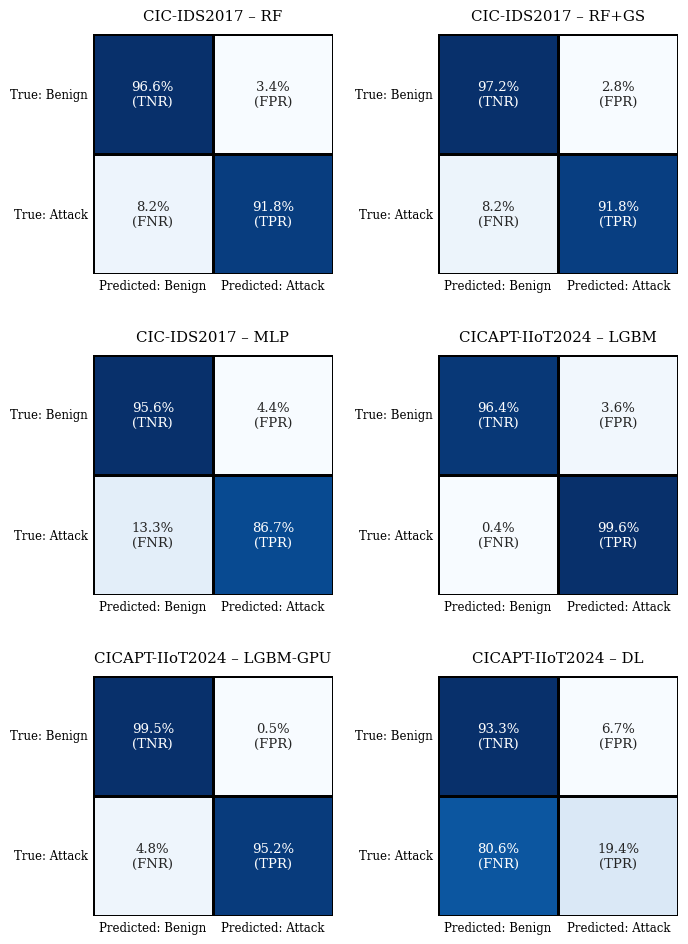}
\vspace{-1.9em}
\caption{Confusion matrices for the evaluated models. %Color intensity is row-normalized to show per-class accuracy (TNR and TPR), while raw counts are overlaid.
}
\label{fig:conf_matrices}
%\vspace{-2em}
\end{figure}

While Table~\ref{tab:key_metrics} provides a high-level summary, the confusion matrices in Figure~\ref{fig:conf_matrices} offer a more granular view of each model's classification behavior. To effectively visualize performance on these highly imbalanced datasets, each $2{\times}2$ confusion matrix is row-normalized so that color intensity directly reflects per-class accuracy: a dark cell in the first row indicates a high \emph{True Negative Rate} (TNR) for the `Benign` class, while a dark cell in the second row denotes a high \emph{True Positive Rate} (Recall) for the `Attack` class. For complete clarity, the absolute number of predictions (TN, FP, FN, TP) are overlaid on each cell. This visual strategy makes it immediately clear, for example, that the \textit{LGBM-GPU} model achieves its leading $F_{2}$-score by significantly reducing False Negatives while keeping False Positives in check, corroborating the quantitative scores in Table~\ref{tab:key_metrics}.

\vspace{-1.1em}
\begin{figure}[htp]
    \centering
    \includegraphics[width=0.9\linewidth]{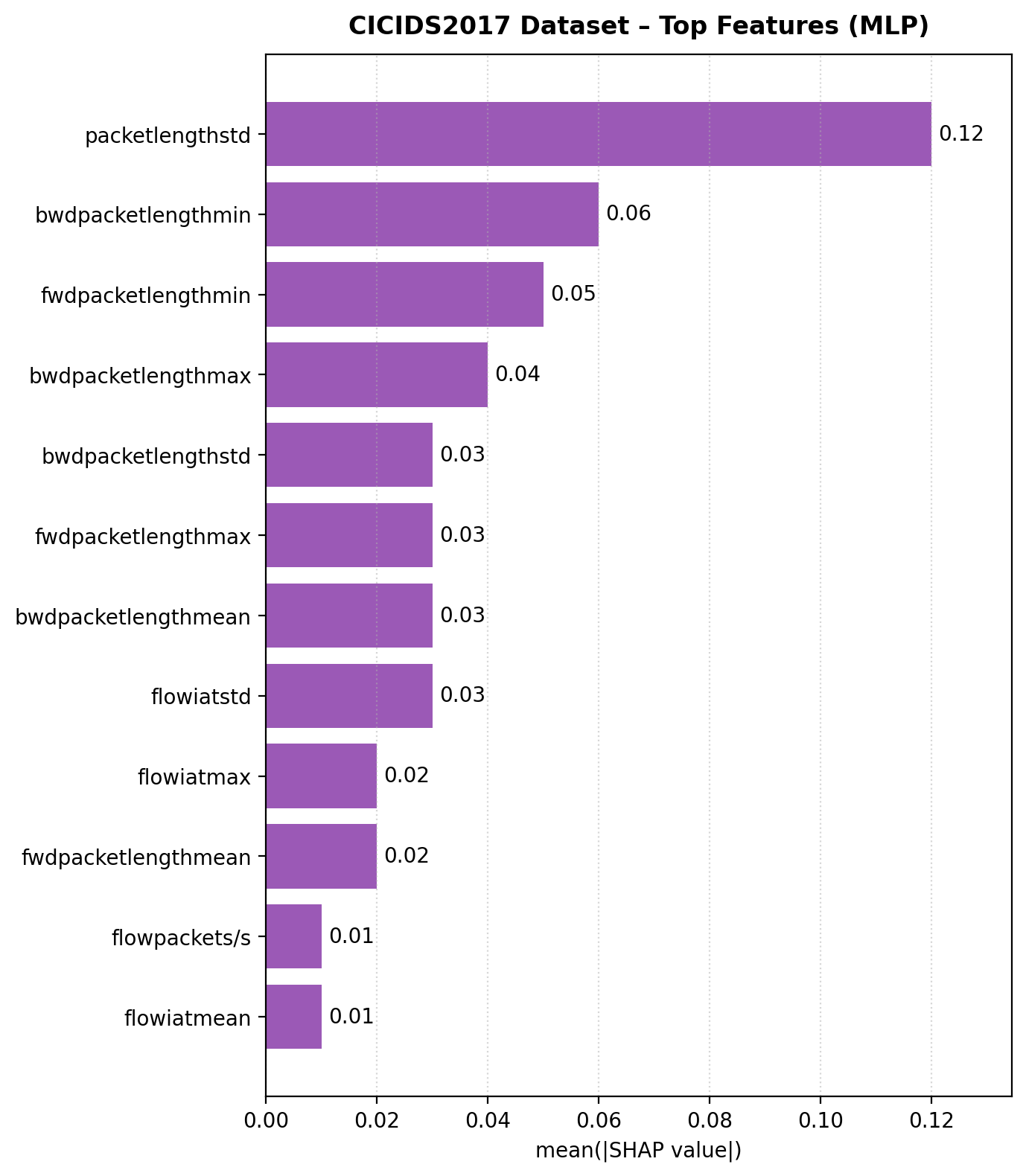}
    \vspace{-0.8em}
    \caption{SHAP feature importance for MLP model  (CIC-IDS2017 dataset).}
    \label{fig:cicids2017_shap}
\end{figure}

As illustrated in Fig.~\ref{fig:cicids2017_shap}, the CIC-IDS2017 dataset is mostly influenced by 
\textbf{packet length–related features} such as \texttt{packetlengthstd}, 
\texttt{bwdpacketlengthmin}, and \texttt{fwdpacketlengthmin}. These metrics highlight abnormal 
packet size variations that are typical of brute-force, DoS, or scanning activities in enterprise 
networks.

\begin{figure}[h]
    \centering
    \includegraphics[width=0.82\linewidth]{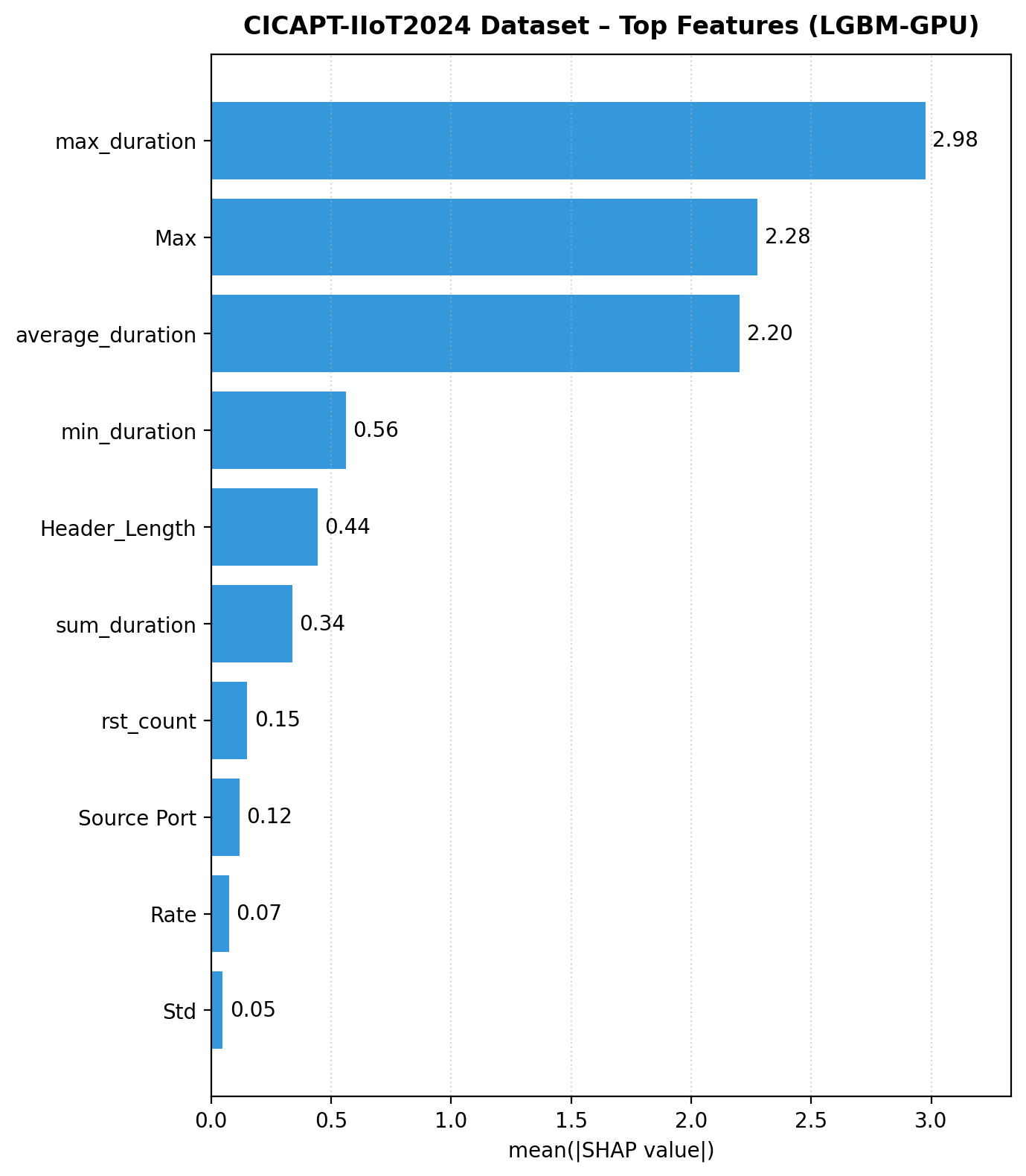}
    \vspace{-0.8em}
    \caption{SHAP feature importance for LGBM-GPU model (CICAPT-IIoT2024 dataset).}
    \vspace{-0.8em}
    \label{fig:cicapt2024_shap}
\end{figure}
%\vspace{-1.0em}

In contrast, Fig.~\ref{fig:cicapt2024_shap} shows that the CICAPT-IIoT2024 dataset is dominated by 
\textbf{duration-based metrics} (\texttt{max\_duration}, \texttt{average\_duration}, 
\texttt{min\_duration}). These features capture the persistence of Advanced Persistent Threats (APTs), 
reflecting their tendency to sustain prolonged or stealthy sessions in IIoT environments. This 
contrast emphasizes that anomaly prediction in IIoT relies more on temporal consistency, whereas 
enterprise networks are driven by packet-based irregularities.

%\vspace{-1em}

%\vspace{-1em}
\section{Conclusion}
This study demonstrates that a digital-twin–based architecture can successfully deliver short-term (T+15 s) anomaly prediction in both conventional enterprise networks (CIC-IDS2017) and Industrial IoT (IIoT) environments (CICAPT-IIoT2024). By combining a time-aware labeling strategy with a recall-oriented F2 metric, the proposed framework significantly improves the early-warning capability of predictive cyber-defense systems. Our experiments show that a GPU-accelerated LightGBM model is particularly effective for this task, achieving a balance of high recall and strong classification performance.

As limitations, the current study treats anomalies only in a binary fashion and addresses class imbalance solely through Random Under-Sampling. 
Future work will explore attack-type–aware evaluation (e.g., DoS, brute force, exfiltration) as well as more advanced balancing strategies such as SMOTE, class-weighting, and ensemble methods to better demonstrate the framework’s versatility and robustness across heterogeneous threat landscapes.
Future work will also formalize twin–plant synchronization by detailing protocol choices and timing/consistency budgets (e.g., REST vs.\ gRPC/MQTT, Kafka backpressure, Redis Streams semantics, and PTP-based clock sync) to strengthen the framework’s communication layer.

In summary, melding time-sensitive prediction models with the programmability of SDN and the modularity of digital-twin designs represents a significant step toward autonomous, proactive cybersecurity. Looking ahead, DT-aided RIS orchestration offers a promising path to further reduce latency and improve reliability in 6G-grade industrial services, aligning our twin’s proactive control with RIS phase-matrix optimization for immersive/metaverse-class applications \cite{masaracchia2024risdt}. The architectural enhancements outlined in this paper pave the way for a future where cyber-physical systems can not only detect ongoing threats but also anticipate and neutralize them before they can cause harm.

%In summary, melding time-sensitive prediction models with SDN programmability and modular digital-twin designs represents a significant step toward \emph{autonomous, proactive} cyber-security.

\section*{Acknowledgment}
M. Erel-Özçevik is also partially supported by The Scientific and Technological Research Council of Turkey (TUBITAK) 1515 Frontier R\&D Laboratories Support Program for BTS Advanced AI Hub: BTS Autonomous Networks and Data Innovation Lab. (Project No:5239903).

\end{document}